\documentclass[sigconf]{acmart}
\AtBeginDocument{%
  \providecommand\BibTeX{{%
    \normalfont B\kern-0.5em{\scshape i\kern-0.25em b}\kern-0.8em\TeX}}}

\copyrightyear{2021}
\acmYear{2021}
\setcopyright{acmcopyright}\acmConference[UMAP '21]{Proceedings of the 29th ACM Conference on User Modeling, Adaptation and Personalization}{June 21--25, 2021}{Utrecht, Netherlands}
\acmBooktitle{Proceedings of the 29th ACM Conference on User Modeling, Adaptation and Personalization (UMAP '21), June 21--25, 2021, Utrecht, Netherlands}
\acmPrice{15.00}
\acmDOI{10.1145/3450613.3456829}
\acmISBN{978-1-4503-8366-0/21/06}
\usepackage{csquotes}

\begin{document}

\title{Privacy as a Planned Behavior: Effects of Situational Factors on Privacy Perceptions and Plans}

\author{A K M Nuhil Mehdy}
\orcid{0000-0003-4820-8380}
\affiliation{%
  \institution{Computer Science Department, Boise State University}
  \city{Boise}
  \state{ID}
  \country{USA}
}
\email{akmnuhilmehdy@u.boisestate.edu }
\author{Michael D. Ekstrand}
\orcid{0000-0003-2467-0108}
\affiliation{%
  \institution{People \& Info. Research Team, Boise State University}
  \city{Boise}
  \state{ID}
  \country{USA}}
\email{michaelekstrand@boisestate.edu}

\author{Bart P. Knijnenburg}
\affiliation{%
  \institution{School of Computing, Clemson University}
  \city{Clemson}
  \state{SC}
  \country{USA}}
\email{bartk@clemson.edu}

\author{Hoda Mehrpouyan}
\affiliation{%
 \institution{Computer Science Department, Boise State University}
 \city{Boise}
 \state{ID}
 \country{USA}}
\email{hodamehrpouyan@boisestate.edu}
\renewcommand{\shortauthors}{Trovato and Tobin, et al.}

\begin{abstract}
To account for privacy perceptions and preferences in user models and develop personalized privacy systems, we need to understand how users make privacy decisions in various contexts. Existing studies of privacy perceptions and behavior focus on overall tendencies toward privacy, but few have examined the context-specific factors in privacy decision making. We conducted a survey on Mechanical Turk (N=401) based on the theory of planned behavior (TPB) to measure the way users' perceptions of privacy factors and intent to disclose information are affected by three situational factors embodied hypothetical scenarios: information type, recipients' role, and trust source. Results showed a positive relationship between subjective norms and perceived behavioral control, and between each of these and situational privacy attitude; all three constructs are significantly positively associated with intent to disclose. These findings also suggest that, situational factors predict participants' privacy decisions through their influence on the TPB constructs.
\end{abstract}





\maketitle

\section{Introduction}
\label{section:intro}
Users' decision to share their personal information, and the perceptions of risk that inform this decision, vary from situation to situation. Situations consist of various factors such as the information type, recipient of the information, and the trust source behind the motivation for sharing. Past research has not paid much attention to how these factors can be used to model and predict users' contextual privacy concerns and decisions. This is an important shortcoming, as decision research suggests that users' privacy preferences are malleable rather than stable and that privacy behavior may vary based on situational and contextual factors \cite{john2011strangers, simonson1992choice, knijnenburg2017privacy}. Moreover, individual's privacy expectations depend on the contexts in which the user is sharing information \cite{mehrpouyan2017measuring,patkos2015privacy,mehdy2020user,joshaghani2019formal}.

In order to understand, model, and possibly predict human privacy behavior in various situated environments, there have been several factors and parameters documented to influence users in their privacy decisions. The theory of planned behavior (TPB) \cite{ajzen1991theory}, an extension of the theory of reasoned action \cite{vallerand1992ajzen}, is a behavioral theory that helps modeling users' perceptions and plans. However, most privacy research based on this theory have either studied single situations, or have considered a very limited set of situational factors \cite{heirman2013predicting, saeri2014predicting}. As a result, understanding the characteristics and impact of various situational factors on users’ privacy decision is still an active area of research. 

In this work, we study users' situational privacy decisions, through a scenario-based survey with 401 participants, each responding to several of 48 different scenarios. Each data point consists of responses to a set of questionnaires that measure participants' attitudinal evaluations of each scenario as well as their perceptions and intention to disclose private information under the specified situation. Alongside the scenario-specific questions, participants responded to a set of general attitude questions to elicit their general attitude towards information disclosure. We perform a path analysis to model participants' privacy perceptions and plans, taking into consideration their attitudinal evaluations on \textit{subjective norm}, \textit{perceived behavioral control}, and \textit{attitude} by manipulating three situational factors: information type, recipient role, and trust source. The results from the analysis reveals how users make privacy decisions in various situations, and how the situational factors have significant effects on users’ perceptions of privacy factors and intention to disclose potentially private information. This paper is the first to our knowledge to combine the Theory of Planned Behavior with a contextual approach to privacy modeling. This study also contributes several insights to the area of user-tailored privacy modeling and personalized privacy systems\cite{knijnenburg2017privacy}, through the following research questions:

\begin{enumerate}
\item How do users' subjective perceptions of TBP constructs differ in different informational situations?
\item How do situational perceptions affect users' intent to disclose information?
\item How do users' situational perceptions and intents relate to their general privacy attitudes?
\end{enumerate}

\section{Background and Related Work}
\label{section:related-works}
Since our path model is based on the theory of planned behavior, we first briefly introduce this behavioral theory in this section. In the following subsections, we review the related research that uses this theory to model users' privacy decision-making process. We also briefly review research that models users' contextual privacy decisions. 

\subsection{Theory of Planned Behavior (TPB)}
\label{section:tpb-background}
According to the TPB, people's behavior is directly determined by their behavioral \textit{intentions}, which are in turn influenced by their \textit{attitude}, perception of the \textit{subjective norms}, and \textit{perceived behavioral control}. Also, the \textit{perceived behavioural control} can, together with \textit{intention}, be used to explain the actual \textit{behavior}. In the literature \cite{beck1991predicting, heirman2013predicting} these constructs are defined as follows:

\begin{itemize}
    \item Attitude (A) is defined by the positive or negative evaluation of the decision (e.g., how well the participant understands the value of an action).
    \item Subjective norm (SN) is defined as a culturally appropriate and desired behavior that is generally expected of a person with in his/her social group  (e.g., how a participant's closest relatives act on similar situation).
    \item Perceived behavioral control (PBC) is defined by the perceived ease or difficulty that the individual addresses to perform the behavior.
\end{itemize}

\begin{figure}[tb]
    \centering
    \includegraphics[width=\linewidth]{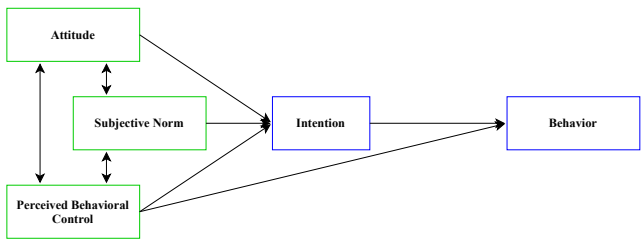}
    \caption{Theory of planned behavior and its core components\cite{ajzen1991theory}.} 
    \Description{Description}
    \label{figure:tpb}
\end{figure}

The theory states that these constructs or components together shape an individual's behavioral intentions. Thus, it provides a model to capture humans' behavioral intention (Figure \ref{figure:tpb}). Theory of planned behavior is used in many research areas and has demonstrated its effectiveness in predicting human behavior in various fields such as privacy\cite{heirman2013predicting, van1999involving}, use of the internet \cite{yao2008predicting}, health \cite{conner2003environmental}, environmental psychology \cite{macovei2015determinants}, etc.

\subsection{Modeling Users' Privacy Decisions Using TPB}
\label{section:tpb-works}
In spite of the dynamic nature of privacy behavior~\cite{john2011strangers, simonson1992choice} and the fact that privacy paradox shows that users' intentions and attitudes may not always result in privacy-protective behaviors~\cite{adjerid2016beyond}, studies have used the TPB to investigate and model the most important factors that influence users' privacy decision-making process~\cite{ajzen1991theory}. Heirman et. al.~\cite{heirman2013predicting} analyzed the impact of the TPB factors (i.e., attitude, subjective norm, perceived behavioral control) on the disclosure of private information through a structured survey. A similar TPB-based approach was utilized by Saeri et. al. to investigate Facebook users' privacy protection tendency based on descriptive norms, risk, and trust \cite{saeri2014predicting}.
Yao et. al. extended the TPB to model users' intention to adopt an online privacy protection strategy~\cite{yao2008predicting}. Their analysis showed that ``the intention to adopt online privacy self-protection is a function of one's attitude towards protective strategies, the subjective norm of adoption, and the perception of behavioral control''. Lwin et. al. combined Laufer and
Wolfe's multidimensional approach to privacy \cite{laufer1977privacy}, and an extended version of Ajzen's theory of planned behavior \cite{ajzen1991theory} to study the privacy behavior of online users \citet{lwin2003model}. They partially used a TPB inspired conceptual framework to investigate the reasons behind users' intention to disguise their identities (i.e., private information). While TPB is normally used for grounding designs and analyses related to any type of human behavior towards an action \cite{van1999involving}, researchers have successfully used TPB for in-depth analysis of privacy attitudes and privacy behavior \cite{ho2017understanding, belkhamza2017effect, burns2013applying} with ample justifications \cite{dienlin2015privacy}.


All of the above-mentioned works have one common limitation: they assumed the users' privacy perceptions (TPB construct measures) to be stable and did not take into account the potential impact of contextual factors. The way contextual dimensions influence TPB remains underexplored.

\subsection{Modeling Users' Contextual Privacy Decisions}
\label{section:contextual-behavior}
Many researchers have studied modeling users' decision-making process in the context of various types and recipients of the information. \citet{knijnenburg2014increasing}, while exploring the design parameters of social network site's privacy-settings UI (user interfaces), discovered about how the type of information and their specific recipients have significant effect on user's sharing tendency. In their study participants were asked to set their privacy settings on a custom made privacy settings UI of an imagined Facebook-like social network site by indicating which of their profile information they would share with whom. At the end of the study, they measured the users’ interpersonal privacy concerns using a post-experimental questionnaire. In another user study, \citet{knijnenburg2013counteracting}  validated the primitive idea of users' privacy calculus (i.e., costs vs benefits which measures the benefits of privacy allowances and the resulting costs \cite{dinev2006extended}) and how it led them to disclose different types of information to different types of websites in a purpose-specific manner. They found that the perceived risk and perceived relevance of the disclosure depends on the interaction between the type of the information and the type of the website/recipient, and that this perceived risk and relevance decreases and increases disclosure, respectively. While both studies show how the perceived relevance and risk of the information---as well as the disclosure activity or intention---depend on the type and recipient of the information, neither of these studies takes into consideration the impact of ephemeral situations (i.e., scenarios or contexts) on the participants' behavior. 

In a contextual setup, \citet{lederer2003wants} investigated the relative effects of information recipient and the situation towards information disclosure. They conducted a study with 130 participants by providing them with two hypothetical situations (working lunch, social evening) and four inquirers/recipients (spouse, employer, stranger, merchant). They asked each participants to imagine using a mobile phone which is capable of collecting and sharing profile and location information to the requesting parties. Through a web based questionnaire they analyzed the user's preferences and found that ``identity of the information inquirer is a stronger determinant of privacy preferences than is the situation in which the information is collected''. However, they found that the situation is also an important determinant but only when the information inquirer is an employer. Even though they incorporated scenarios and recipients' role in the study, the characteristics of the scenarios were unchanged and represents only two static situations. In this regard, their contextual behavior analysis is limited to these two situations only. Nevertheless, the above-mentioned work and other similar works have demonstrated the influence of various contextual factors on users' privacy behavior \cite{patil2005gets, olson2005study}.

\subsection{Representing Contexts with Scenarios}
\label{section:context-scenario}
One way of contextualizing a survey is to introduce various scenarios to the participants and ask them to respond to questionnaires linked to each of those scenarios \cite{lederer2003wants}. However, one challenge in this regard is to create proper scenarios with an appropriate level of detail. Researchers from the area of scenario-based survey have introduced many different approaches to create hypothetical scenarios using text, graphics, games, app interfaces, etc. \cite{emami2018influence, kumar2017no, wijesekera2018contextualizing, zhao2019make}. Among all of these, text-based scenarios are preferred in case of surveying the participants. A set of methods have been well-established for the development of such scenarios, especially in the privacy survey domain, such as the factorial method, storytelling method, and claim analysis. The factorial method involves creating scenarios ``based on a set of predefined factors that describe all or a subset of possible combinations seen in a situation or decision problem'' \cite{brauer2009creating}. These factors could be socioeconomic, behavioural, or clinical issues, defined as categorical variables with two or more levels. However, the number of factors and their levels are subject to be decided carefully. Otherwise, the number of combinations of factor categories increases very rapidly, which in turn increases the total number of unique scenarios. On the other hand, the storytelling method suggests creating a few illustrative scenarios, usually based on the experience of the members of the research team. In our work, we adopt the former method to create the scenarios while keeping the number of factors and their categories low. 

\section{Survey Methodology}
\label{section:methodology}
\begin{figure}[tb]
    \centering
    \includegraphics[width=\linewidth]{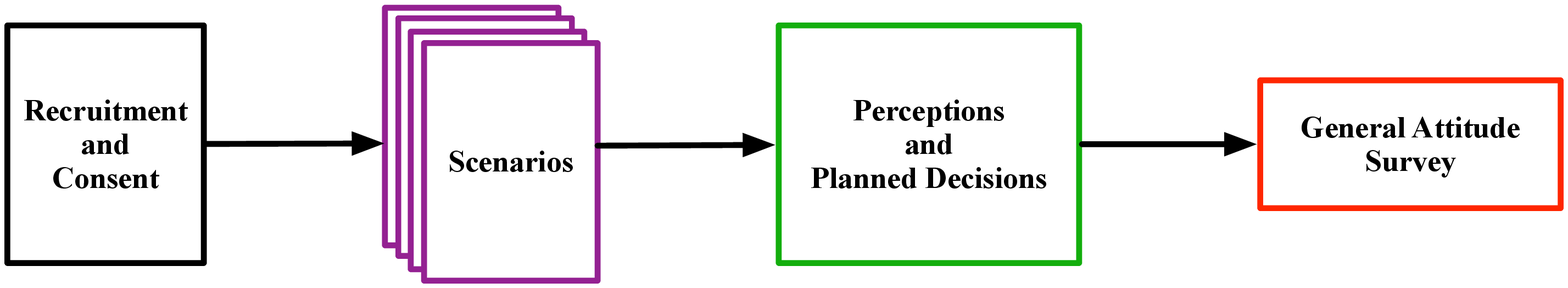}
    \caption{Overview of the experimental flow.} 
    \Description{Description}
    \label{figure:exp-flow}
\end{figure}

The overall flow of our experiment can be divided into three main steps: i) recruitment and consent ii) capturing scenario-specific perceptions and planned decisions iii) general attitude survey (Figure \ref{figure:exp-flow}). After consenting to the study, a participant is assigned a set of 8 random hypothetical scenarios and asked to respond to those scenarios one after another. Each scenario gives the participant a situation in which he/she must decide whether or not to share a piece of information. This incorporates the situational factors on which participants might have a degree of reliance for their perception and decision towards disclosure intention (see Section~\ref{subsection:manipulation}). A participant has to read a given scenario and respond to all of the corresponding questions before proceeding to the next assigned scenario.

In the next step, the participant takes a short survey to capture their general privacy attitudes independent of any particular scenario. This step is designed to capture such perceptions that are assumed to be stable over time and do not usually change based on any situation. There is another final step for collecting the demographic information of the participants, in which participants are asked to optionally input their gender, age group, country of residence, and the duration of residence in that country. It is worth mentioning that the presented scenarios are hypothetical; none of the participants' personally identifiable information is collected in any step of the survey, explicitly or implicitly.

\subsection{Factor Manipulation}
\label{subsection:manipulation}
We manipulate 3 situational factors in order to measure their effect on participant response:

\begin{description}
\item[Information Type (IT)]
    The general category information that may be disclosed. Each scenario is about one of three categories: health, finance, or relationship.
    
\item[Recipient's Role (RR)]
    The kind of recipient the information may be disclosed to, with their relationship to the participant.
    We use four roles: family, friend, colleague, and online (e.g. discussion forum).

\item[Trust Source (TS)]
    Where the idea of disclosing this information to this recipient came from.
    We test four trust sources: family, friend, expert (e.g. physician or financial adviser), and online (e.g. searching the web).
\end{description}

This choice of factors is partly inspired from the theory of contextual integrity (CI) \cite{nissenbaum2004privacy, barth2006privacy}. The CI theory provides the ground of informational norm where, norm is formulated as a tuple of access permission $(\rho, \tau)$, environmental conditions ($\psi$), and transmission principle ($\eta$). Hence, a norm, $n$ is represented as: $n =((\rho, \tau), \psi, \eta)$ where, $n$ = Informational norm, $\rho$ = Recipient's Role, $\tau$ = Information type, $\eta$ = Transmission principle. These factors yield a total of 48 (3*4*4) unique situations. Every situation and the associated questionnaire is intended to measure the situational privacy perceptions of the participant through 3 constructs: i) attitude ii) subjective norm iii) perceived behavioral control

\subsection{Scenario Generation}
\label{section:scenario-development}
For each combination of situational factors, we wrote a scenario in which a trust source encourages the participant to share information with a recipient.
To minimize extraneous variability, we made each scenario as similar as possible while presenting the combination of factors in a natural and coherent manner.
As an example, the scenario for \textit{health} as information type, \textit{friend} as trust source, and \textit{family member} as recipient's role is:
\begin{displayquote}
Your doctor called and told you that your lab results came back positive for a disease. One of your friends suggested discussing the situation with a family member and asking their support, saying it could be helpful.
\end{displayquote}
Changing the trust source from friend to family and recipient's role from family to online yields another scenario: 
\begin{displayquote}
Your doctor called and told you that your lab results came back positive for a disease. A family member suggested asking other patients and doctors on an online discussion forum, saying they have found it helpful for dealing with their similar condition.
\end{displayquote}

In this study, the domains of the scenarios are health, finance, and relationship. This means, we have generated 3 sets of scenarios for these three types of information. Each of these sets contains 16 different scenarios (i.e., 4 RR x 4 TS values) resulting in a total of 48 scenarios. For each scenario, the participants answered a set of questions to measure their perception of TPB constructs in that scenario and indicated whether or not they would share the information. 

\subsection{Scenario Randomization}
\label{section:scenario-randomization}
As discussed earlier, every participant is assigned a set of 8 random scenarios with associated questionnaires. To ensure a minimum level of variability within each user's situations (and therefore responses), we used rejection sampling to require that each user's 8 scenarios covered all 11 distinct factor levels at least once.
Redrawing a fresh, independent set of 8 scenarios if a user's initial assignment excludes a level ensures maximal statistical independence subject to our inclusion requirement. We further randomly order scenarios for each participant to avoid order effects. Also, we implicitly account for the variability of judgements of the questions and scales across the participants by setting random per-user intercepts while doing the analysis.

\subsection{Testing the Experiment}
\label{section:pilot-test}
We piloted the experiment and surveys with 6 colleagues from our research lab.  Their feedback helped fix issues in the survey application, user-experience/user-interface, and clarity of the scenarios and questions.  We then soft-launched the survey on Amazon Mechanical Turk with an initial round of 10 participants to collect information on the average time needed to complete the survey and estimate total survey cost.

\subsection{Participants}
\label{section:participants}

We recruited the participants for the final survey via Amazon Mechanical Turk, an online crowd-sourcing marketplace.  We filtered for Workers from the USA with a good reputation (i.e., at least 95\% HIT approval rate and 50 hits approved) who are at least 18 years old. We paid \$2.00 per survey based on pilot trials indicating Turkers could complete it in about 15 minutes.

\subsection{Data Collection and Cleaning}
\label{section: data-collection}
We employed a number of filters to ensure the quality of the data. First of all, we capture the time a participant spent on each scenario step and removed the data points (i.e., responses associated with a specific scenario) from our analysis if the spent time was too low (less than 15 seconds per scenario) to be realistic. Secondly, we embedded attention check questions randomly in between survey questions on two surveys per participant, and removed 9 data points for failing the attention check. Since participation is anonymous and therefore a participant could potentially submit several responses, we restricted this incident by setting a browser cookie for 3 days after a successful submission.

We converted the 5-point scale responses to TPB questions (ranging from \textit{Strongly Disagree} to \textit{Strongly Agree}) into a numeric format (1 to 5). We represent the \textit{Share} and \textit{Not Share} options for the final decision question in logical numeric form, 1 and 0. We dummy-coded categorical variables for the situational factors.  We then computed a standardized scale-score for each TPB construct by taking the mean of the responses on its questions (see Section \ref{section:questionnaire}), after inverting negative questions, so that 5 represents the opinion most in favor of sharing for each question.

\section{TPB-Based Questionnaire and Path Model}
\label{section:path-model}
\begin{figure}[tp]
    \centering
    \includegraphics[width=\linewidth]{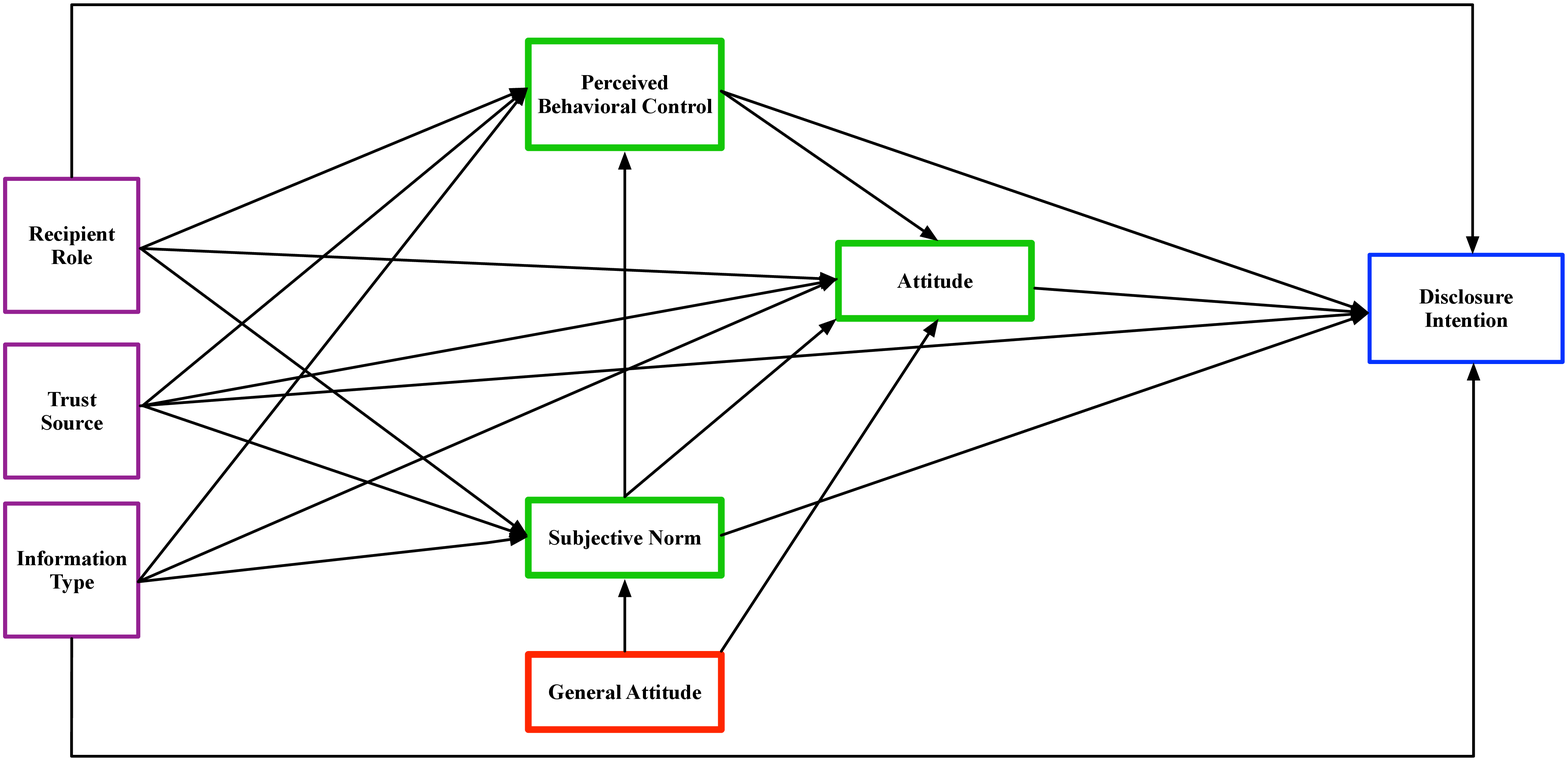}
    \caption{The initial path model.}
    \Description{Description}
    \label{figure:path-model-initial}
\end{figure}

As previewed in Section \ref{section:tpb-background}, we designed our survey to measure participants' behavioral intention and their situational perception of three constructs from TBP: attitude (A), subjective norm (SN), and perceived behavioral control (PBC).
We followed the scenario-specific questionnaires with a short survey to assess participants' general attitude towards privacy.
We integrated the TPB constructs, manipulated factors, and general privacy attitude into an initial path model shown in Figure \ref{figure:path-model-initial}. The colors on the figure follow the convention of Knijnenburg et al.'s evaluation framework \cite{knijnenburg2012explaining}, where purple = manipulations, green = subjective evaluations, red = personal characteristics, and blue = behavior.
We evaluate this path model through a causal modeling technique called \emph{path analysis} to determine if our causal model fits the survey data well. Note that path analysis is ``not intended to discover causes but to shed light on the tenability of the causal models that a researcher formulates'' \cite{pedhazur1997multiple}. We apply this technique to examine the relationships between the observed variables in terms of the strength and direction of the path beta coefficients.

\subsection{Model Specification}
The dummy variables representing the three scenario parameters---information type, recipient's role, and trust source---comprised the \textit{exogenous} variables(variables that have arrows outbound from them and not caused by any other variables of the model \cite{streiner2005finding}) in the preliminary path model, together with general attitude. Trust source was eliminated from the final model because of its non-significant association with the TPB constructs. In our initial model the exogenous variables were causally related to attitude, perceived behavioral control, and subjective norm, although some of these relations (e.g. from information type to perceived behavioral control) were removed due to a lack of significance. Relationships among attitude, perceived behavioral control, and subjective norm were also modeled. Finally, all variables were causally related to disclosure intention, although only attitude, perceived behavioral control, and subjective norm were found to be significant. The final model has a total of 27 free parameters, and 28 fixed parameters whose values are estimated from the data.

We fit the model with Mplus, a statistical analysis tool for conducting the analysis as well as constructing the diagram of our path model \cite{muthen1998mplus}.

\subsection{Questionnaire}
\label{section:questionnaire}
The survey contains two sets of questions - \textit{i) scenario specific questions (12 in total) ii) general attitude questions (4 in total)}. The first set of 12 questions are repeated for each of the 8 assigned scenarios to each participant. The second set of questions are presented at the last step of the survey. The scenario-specific questionnaire is inspired by \citet{heirman2013predicting}, which operationalized the constructs in the theory of planned behavior \cite{ajzen1991theory}. The second set of questions is inspired by prominent privacy research \cite{buchanan2007development, ackerman1999privacy}. 

The following questions were asked once per scenario:
\begin{enumerate}
\item \textit{Attitude (Cronbach's alpha: 0.68)}
\begin{enumerate}
\item I would benefit from sharing this situation. (Scale: Completely disagree (1) to Completely agree (5))
\item I am concerned about where this information would be stored or recorded if I shared it with \textit{\$recipient}.  (1-5, reversed)
\item I do not expect any significant risks if I share this situation. (1-5)
\item I have concerns about who will learn about this situation. (1-5, reversed)
\end{enumerate}

\item \textit{Subjective Norm (Cronbach's alpha: 0.79)}
\begin{enumerate}
\item I think my friends or family would share in this situation. (1-5)
\item A friend or family member would likely suggest that I disclose this situation. (1-5)
\item My friends would approve of me disclosing this situation. (1-5)
\item Some people in my life would disapprove if they knew I shared this situation. (1-5, removed from the scale)
\end{enumerate}

\item \textit{Perceived Behavioral Control (Cronbach's alpha: 0.66)}
\begin{enumerate}
\item I have control over how my information will be used after I share it in this situation. (1-5)
\item I trust the recipient of my information to honor my wishes if I ask them to keep my situation a secret. (1-5, removed)
\item Sharing this situation would put me at risk. (1-5)
\end{enumerate}

\item \textit{Disclosure Intention}
\begin{enumerate}
\item What would you do in this scenario? (Scale: Not share (0) or Share (1))
\end{enumerate}
\end{enumerate}


The following questions were asked once per participant:
\begin{enumerate}
\item \textit{General Attitude (Cronbach's alpha: 0.68)}
\begin{enumerate}
\item In general, I am concerned about threats to my personal privacy. (1-5, reversed)
\item I am generally concerned about my privacy while using the internet. (1-5, reversed)
\item I believe other people are too concerned about online privacy issues. (1-5, removed)
\item I think I am more sensitive than others about the way my contacts handle information I consider private. (1-5, reversed)
\end{enumerate}
\end{enumerate}

We performed Cronbach's alpha test to measure the items' scale reliability. Thus, item (d) was removed from the subjective norm scale because of its negative effect on the alpha score. We removed item (c) from perceived behavioral control, and item (c) from general attitude because of the same reason. Items (b) and (d) in the attitude scale were reversed while calculating their score because of their negative phrasing. All items in the general attitude scale were reversed to align this factor with the context-specific attitude.

\section{Results}
\label{section:results}
\begin{figure*}[tb]
    \centering
    \includegraphics[width=\textwidth,height=\textheight,keepaspectratio]{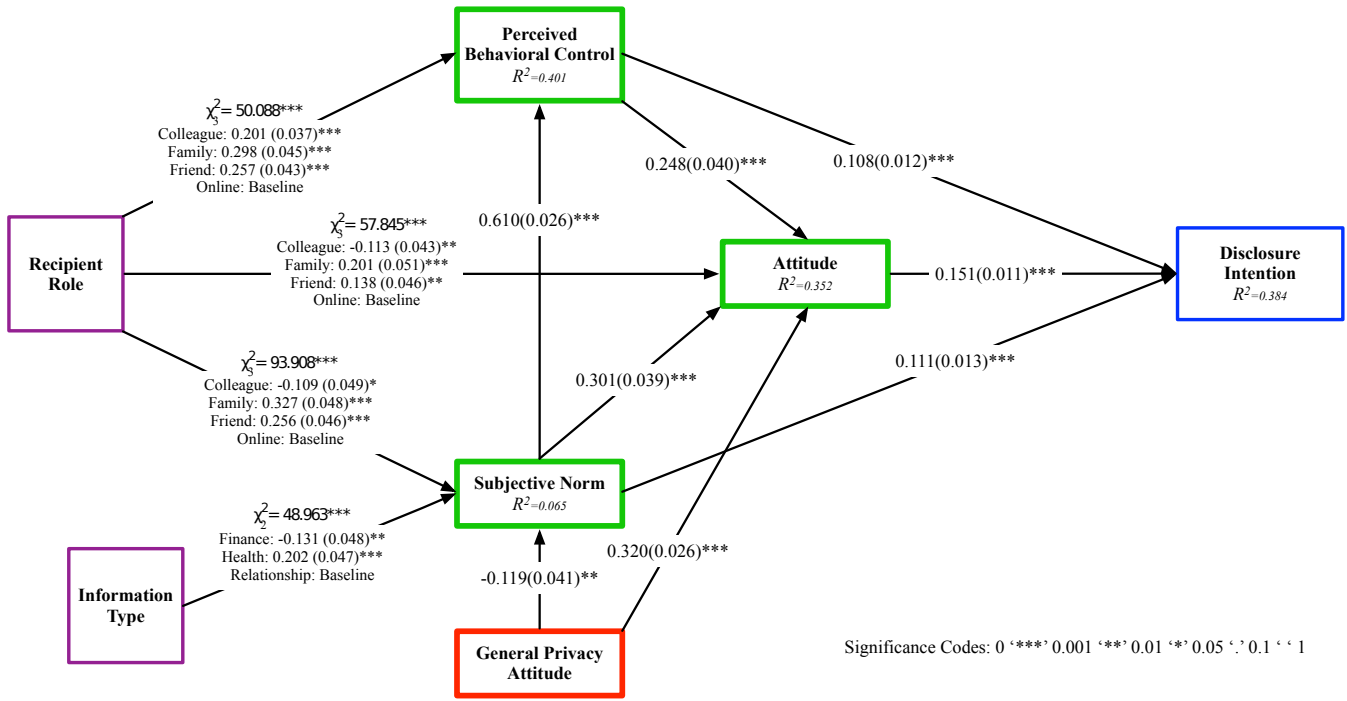}
    \caption{Path model results. Paths that are non-significant ($p > .05$) are removed from the model.}
    \Description{Description}
    \label{figure:path-model}
\end{figure*}

This section describes the path analysis results in detail. First we talk about the descriptive analysis and the quality of the model fit. Then we describe the direct and indirect effects of the factors and constructs in subsequent sections. Figure \ref{figure:path-model} depicts our final path model.

\subsection{Descriptive Statistics}
\begin{figure}[h]
  \centering
  \includegraphics[width=\linewidth]{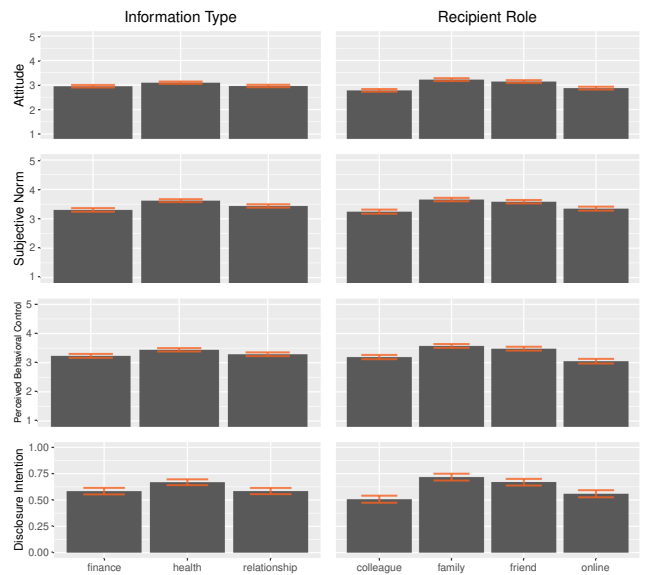}
  \caption{Constructs vs Mean Scale-score based on Information Type and Recipient's Role.}
  \Description{Description}
  \label{figure:bars}
\end{figure}

Table~\ref{tab:demo} reports the demographic information of the participants. We share this information not because these are relevant factors in this context but for those who may attempt to reproduce this results with a similar setup. Figure \ref{figure:bars} reports the differences in attitude, subjective norm, perceived behavioral control, and disclosure intention between the different value of the scenario parameters ``information type'' and ``recipient role'', including standard error bars. For example, we can see how the participants perceive a higher level of behavioral control when the recipient is family member or friend than that of colleague or online platforms.

\begin{table}[h]
  \caption{Demographic information of the participants.}
  \label{tab:demo}
  \begin{tabular}{ll}
    \toprule
    \textbf{Constructs}&\textbf{Distribution}\\
    \midrule
    Gender & Man: 252\\
     & Woman: 144\\
     & Not Answered: 3\\
     & Non Binary: 1\\
     & Woman,Man: 1\\
    \midrule
    Age & 18-30: 108\\
     & 31-40: 148\\
     & 41-50: 75\\
     & 51-60: 49\\
     & 60+: 19\\
     & Not Answered: 2\\
  \bottomrule
\end{tabular}
\end{table}

\subsection{Model Fit}
Figure \ref{figure:path-model} depicts the final results of the path model analysis in detail. The model fits the data very well with $\chi^2_{11}=12.017$, $p=0.3623$, $CFI=1.0$, $TLI=0.99$, $SRMR=0.008$, $RMSEA=0.005$, $90\%$  $CI=0.000$ to $0.020$. A non-significant $\chi^2$ value $(p > .05)$ is indicative of a path model that fits the data well \cite{shah2012multivariate}. Also, the comparative fit index (CFI) and Tucker-Lewis index (TLI) values which ranges from 0 to 1 show near-perfect scores. Moreover, the relationships in the model explain 38.4\% ($R^2=0.384$) of the variance in disclosure intention, 35.2\% of the variance in attitude, 6.5\% of the variance in subjective norm, and 40.1\% of the variance in perceived behavioral control. 


\subsection{Effect of the Scenario Parameters on TPB Constructs}
\label{section:param-construct-effects}
This section describes the significant effects of the scenario parameters (recipient role and information type) on the TPB constructs (the privacy perceptions of the user). These effects are measured by the paths from the purple (square) boxes to the green (rectangular) ones in Figure \ref{figure:path-model}.

\begin{enumerate}
    \item The recipient's role in the scenario has a significant influence on perception of behavioral control. Compared to "people online", participants are estimated to perceive significantly more control when the recipient is a colleague (0.201 SD higher), a family member (0.298 SD higher) or a friend (0.257 SD higher).
    \item Likewise, the recipient's role in the scenario has a significant influence on attitude. Compared to "people online", people are estimated to have significantly more positive attitude toward disclosure when the recipient is a family member (0.201 SD higher) or a friend (0.138 SD higher), but more negative attitude when the recipient is a colleague (0.113 SD lower).
    \item The recipient's role in the scenario has a significant influence on subjective norm. Compared to "people online", participants are estimated to believe that individuals close to them would be more likely to agree with the scenario when the recipient is a family member (0.327 SD higher) or a friend (0.256 SD higher), but less when the recipient is a colleague (0.109 SD lower).
    \item The information type in the scenario has a significant influence on subjective norm. Compared to "relationship information", participants are estimated to believe that individuals close to them would be more likely to agree with the scenario when the information type is health (0.202 SD higher) but less reliance when the information type is finance (0.131 SD lower).
\end{enumerate}
    
\subsection{Effects between General Attitude and Situational Perceptions}
\label{section:ga-constructs-effects}
We now turn to the relationships between constructs, both situational TPB constructs and the influence of general attitude on these constructs.
The following effects refer to the paths among the green (rectangular) boxes and between the red (rectangular) box and the green ones in Figure \ref{figure:path-model}. 

\begin{enumerate}
    \item The participants' perceived subjective norm regarding the scenario is positively associated with their perception of behavioral control. A 1 SD difference in subjective norm results in an estimated 0.610 SD difference in perceived behavioral control.
    \item Participants' subjective norm is also positively associated with their attitude towards disclosure. A 1 SD difference in subjective norm results in an estimated 0.301 SD difference in attitude.
    \item The perception of behavioral control of the participants regarding the scenario is positively associated with their attitude towards disclosure. A 1 SD difference in perceived behavioral control results in an estimated 0.248 SD difference in attitude.
    \item The participants' general attitude is positively associated with their situational attitude towards disclosure. A 1 SD difference in general attitude results in an estimated 0.320 SD difference in attitude.
    \item General attitude is also negatively associated with perceived situational subjective norm. A 1 SD difference in general attitude results in an estimated -0.119 SD difference in perceived subjective norm.
\end{enumerate}

\subsection{Effects of Situational Perceptions on Disclosure Intention}
\label{section:construct-di-effects}
This section briefly describes about the significant effects between the situational TPB constructs (the privacy perceptions of the user) and users' disclosure intention. The following  effects refer to the paths between the green (rectangular) boxes and the blue (rectangular) one in Figure \ref{figure:path-model}. 

\begin{enumerate}
    \item Participants who perceived a higher level of behavioral control were more likely to engage in the disclosure described in the scenario. Particularly, the odds of disclosure of participants who have a 1 SD higher level of perceived behavioral control are estimated to be 11.4\% higher.
    \item Participants who have a higher level of perceived subjective norm were more likely to engage in the disclosure described in the scenario. Particularly, the odds of disclosure of participants who have a 1 SD higher level of perceived subjective norm are estimated to be 11.7\% higher.
    \item Participants who have a more positive attitude were more likely to engage in the disclosure described in the scenario. Particularly, the odds of disclosure of participants who have a 1 SD higher level of attitude are estimated to be 16.2\% higher.
\end{enumerate}

Although not directly comparable, it's worth mentioning a comparison with the results from \cite{heirman2013predicting} in this section while showing the relationships between the TPB constructs and disclosure intention.  According to their analyses which take into account only the stable factors, an individual's intent to disclose is influenced primarily by a subjective norm and subsequently by attitude, not significantly by perceived behavior control. In contrast, our study shows the order of significant influence of the TPB constructs to disclosure intention as, attitude $>$ subjective norm $>$ perceived behavioral control. It should be noted that in our study, the TPB constructs are already affected by the situational factors.

\subsection{Total Effects of the Scenario Parameter on Disclosure Intention}
\label{section:param-di-effects}

All effects of scenario parameters on disclosure intention were fully mediated by perception of TPB constructs---that is, after controlling for scenario effects through TPB constructs, there were no statistically significant residual effects of scenario parameters on disclosure intention.
This section describes the total significant (indirect) effects of the scenario parameters on the users' disclosure intention. The following effects do not refer to any direct paths between the purple (square) boxes and the blue (rectangular) one in Figure \ref{figure:path-model}. Rather, they refer to the paths from the leftmost boxes to the rightmost box via the mediator rectangular boxes in between.
These total effects describe \textit{how} users' intention changes from one scenario to another; the mediating TBP factors provide an explanation for \textit{why}. The latter may help with future generalizability.

\begin{enumerate}
    \item With regard to the recipient's role in the scenario, compared to the recipient ``people online'', the odds of disclosure were estimated to be 16.6\% higher when the recipient was a family member and 12.9\% higher when the recipient was a friend. Both of these differences were significant ($p = 0.000$ and $p = 0.000$, respectively). There was no significant difference between the recipient ``people online'' and a colleague.

    \item With regard to the type of information, compared to relationship information, the odds of disclosure were estimated to be 3.1\% lower when the scenario involved financial information and recipient was a family member and 5.1\% higher when the scenario involved health information. Both of these differences were significant ($p = 0.007$ and $p = 0.000$, respectively).
\end{enumerate}

\section{Discussion}
\label{section:discussion}
The results from our path analysis show how users make privacy decisions in various situations: the situational factors have significant effects on users' perceptions of privacy factors, which in turn have an effect on their intention to disclose their private information. Unlike most existing studies of privacy perceptions and behavior modeling, we developed a set of unique scenarios by manipulating parameters to imitate various situations and used a TPB-based model to introduce mediating factors that explain the effects of these situational factors on participants' disclosure intentions. This situation-specific extension of the TPB fulfils our initial goal of understanding users' contextual privacy decision-making process.

This study reveals that the recipient’s role in the scenario has a significant influence on peoples' perception of behavioral control, their attitude, and subjective norm (RQ1). People are estimated to perceive a higher level of behavioral control when the recipient is a family member, a friend, or a colleague than when the recipient is people online (e.g., social media, forum etc.). Likewise, people are estimated to have a more positive attitude toward disclosure when the recipient is a family member or a friend than people online, but a less positive attitude when the recipient is a colleague. Users' subjective norm also shows similar order of perceptions. As a result of these effects, people are more likely to disclose their information to friends and family than to colleagues or people online.

The information type in the scenario also has significant influence on participants’ subjective norm. The model shows that people believe that individuals close to them would be most likely to agree with the scenario when it involves health information, followed by relationship information, and finally financial information. These differences propagate to small differences in disclosure intentions as well.

The results from the analysis also show that participants’ perceived subjective norm regarding the scenario is positively associated with their perception of behavioral control and attitude towards intention to disclose (RQ2). In other words, one can make a hypotheses that when users perceive an expectation to share, they also expect that sharing to be respected? Likewise, their perception of behavioral control is a good predictor of their attitude. Moreover, from the results, we can see the positive effects of these three constructs on users' disclosure intention. Users' attitude has the strongest effects on their disclosure intention relative to the other two constructs. Participants with a higher level of positive attitude were more likely to engage in the disclosure described in the scenario. Section \ref{section:construct-di-effects} contains specific detail of these effects. Additionally, our results reveal the significant influence of general attitude on some TPB constructs (RQ3). Participants' general attitude is positively associated with their situational attitude towards disclosure. In contrast, general attitude is negatively associated with perceived situational subjective norm.

Most importantly, our study demonstrates that the effects of the contextual parameters (the recipient's role and information type) on the users' disclosure intention was fully mediated by participants’ attitude, subjective norm, and perceived behavioral control. As such, these TPB constructs serve as significant and sufficient mediators explaining why users disclose more information in some scenarios than in others. These findings contribute important insights to the area of user-tailored privacy modeling and personalized privacy systems by providing a quantitative analysis of the privacy decision making factors.

\subsection{Limitations}
Even though path analysis is often referred to as a causal inference technique\cite{barbeau2019path}, readers should be advised that this model reveals the predictive properties between the factors and constructs. These properties are measured in terms of path coefficients. Therefore, our path analysis shows how the hypothesized model fits the survey data which in turns aims to explain users' privacy decision making process. We also acknowledge that we are only manipulating a few levels per factor in our study, and there could be much more granularity in the information type, recipient's role, and trust source factor; future work should explore this. Additionally, since the results reveal significant relationships between situational factors and disclosure intention, we feel the necessity to integrate additional factors in future studies.

We also note that our scenarios had a hypothetical nature, and hence did not measure actual disclosure but rather users' \emph{intention} to disclose their private information. This is a limitation that our work shares with many other privacy studies \cite{yao2008predicting}, especially in light of the ``privacy paradox'' which shows a discrepancy between disclosure intentions and behaviors, as behaviors tend to be influenced by extraneous factors like default settings and choice framing \cite{anaraky2020exacerbating}. Arguably, though, the absence of such extraneous influences makes users' disclosure intentions a more honest representation of their privacy preferences.

\section{Conclusion}
\label{section:conclusion}
In this paper we have presented the results of a scenario-based survey to understand users' \emph{situational} privacy perceptions and disclosure intentions. These results constitute a contextualized understanding of users' privacy behaviors, connected to the Theory of Planned Behavior, and provide new insights that can help build future user-tailored privacy models. The impact of various situational factors on users' privacy decision is still an active area of research; one particular need is more study of the gap between users' intention versus reported and actual behavior. In future work, we plan to bridge the gap between intention and behavior by incorporating reported or actual behavior in the model. We also plan to evaluate the predictive power of the current path analysis by surveying a new sample of users. Moreover, we plan to increase the sample size significantly and employ machine learning based algorithms along with the statistical approaches, as a means to compare various analysis methods for explaining contextual privacy behavior. For now, we can advise the user-modeling community to take the recipient and information type into account when modeling users' situation-specific privacy concerns, and to perhaps build these models not as a uni-dimensional construct, but to include aspects of behavioral control, social norms, and attitude, as suggested by the Theory of Planned Behavior.

\begin{acks}
The authors would like to thank National Science Foundation for
its support through the Computer and Information Science and
Engineering (CISE) program and Research Initiation Initiative(CRII)
grant number 1657774 of the Secure and Trustworthy Cyberspace
(SaTC) program: A System for Privacy Management in Ubiquitous
Environments.
\end{acks}

\bibliographystyle{ACM-Reference-Format}
\bibliography{main}

\appendix









\end{document}